\documentclass[unsortedaddress,superscriptaddress,prl,twocolumn]{revtex4-2}
\usepackage{hyperref}
\usepackage{epsfig}
\usepackage{graphicx}
\usepackage{subfigure}
\usepackage{latexsym}
\usepackage{color}
\usepackage{fullpage}
\usepackage{dcolumn}
\usepackage{bm}
\usepackage[normalem]{ulem}
\usepackage{units}
\usepackage{amsmath}
\usepackage[paperwidth=210mm,paperheight=297mm,centering,hmargin=2cm,vmargin=2.3cm]{geometry}
\usepackage{eqnarray,amsmath}

\begin{document}

\title{Two-way twisting of a confined monolayer: orientational ordering within the van der Waals gap between graphene and its crystalline substrate}

\author{Simone Lisi}
\author{Val\'{e}rie Guisset}
\author{Philippe David}
\author{Estelle Mazaleyrat}
\author{Ana Cristina G\'{o}mez Herrero}
\author{Johann Coraux}
\email{johann.coraux@neel.cnrs.fr}
\affiliation{Universit\'{e} Grenoble Alpes, CNRS, Institut NEEL, Grenoble INP, 38000 Grenoble, France}

\begin{abstract}
Two-dimensional confinement of lattices produces a variety of order and disorder phenomena. When the confining walls have atomic granularity, unique structural phases are expected, of relevance in nanotribology, porous materials or intercalation compounds where \textit{e.g.} electronic states can emerge accordingly. The interlayer's own order is frustrated by the competing interactions exerted by the two confining surfaces. We revisit the concept of orientational ordering, introduced by Novaco and McTague to describe the twist of incommensurate monolayers on crystalline surfaces. We predict a two-way twist of the monolayer as its density increases. We discover such a behavior in alkali atom monolayers (sodium, cesium) confined between graphene and an iridium surface, using scanning tunneling microscopy and electron diffraction.

\end{abstract}

\maketitle

\textit{\textbf{Introduction. -- }}In two dimensions, the competition between distinct kinds of orders and the rich related physical phenomena can be controlled within stacks of atomic layers individually strained or rotated. This was resoundingly illustrated with bilayers of two-dimensional (2D) materials such as graphene and transition metal dichalcogenides, where strong electron correlation effects were found to emerge for specific twist (relative rotation) angles and corresponding moir\'{e} lattices \cite{cao,li}. Now, twisted trilayers of 2D materials offer new playgrounds, where three instead of two kinds of order compete, thereby expanding the range of possibilities to tune physical properties via structure engineering \cite{zhu,ramires,qin,park}.

Beyond these artificially-made systems, others, based on 2D materials epitaxially grown on crystalline substrates, also feature competing orders that modify and enrich the properties of the 2D material, for instance with replica bands \cite{pletikosic}, van Hove singularities \cite{brihuega} and flat electronic bands \cite{ehlen}, controlled by coexisting periodicities, twist angle or moir\'{e} lattices. These epitaxial systems can be intercalated with monolayers of various elements \cite{daukiya}, meaning that there also, three competing orders can coexist and a rich phenomenology of physical effects is to be expected.

This is the configuration we address here, under a structural viewpoint. We revisit seminal works by Novaco and McTague from 1977, back then focusing on bilayer systems. They predicted, even for a layer made of loosely-bond units (atoms, molecules), incommensurate with their substrate, a global ordering  with a continuous and monotonous variation of the twist angle $\phi$ \textit{versus} the atomic density $\rho$. What they coined `orientational ordering' \cite{novaco,mctague} was later observed experimentally \cite{shaw,wang,meissner,jin}. We introduce a modified Novaco-McTague model, adapted to the situation of a monolayer confined between two rigid crystalline surfaces. We predict a unique kind of orientational ordering, not monotonous anymore but instead featuring a two-way twist between configurations, reached at specific $\rho$ values, where the monolayer locks onto the bottom and top lattices. Next, using scanning tunneling microscopy (STM) and reflection high-energy electron diffraction (RHEED), we present a practical realisation of this new flavour of orientational epitaxy, in monolayers of alkali atoms (Na, Cs) of varying density, sandwiched between graphene and Ir(111).

\begin{figure*}[!ht]
\begin{center}
\includegraphics[width=16.4cm]{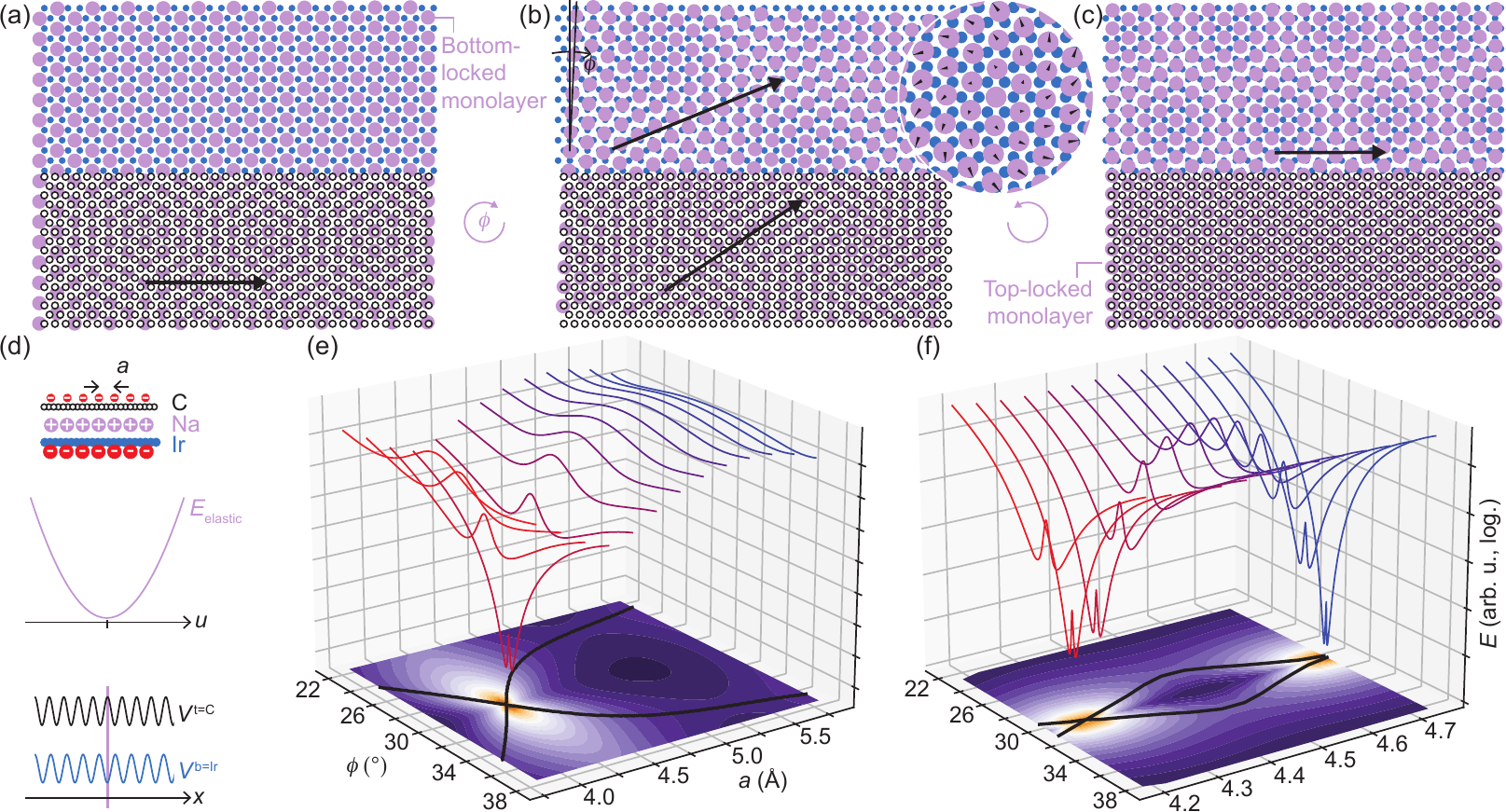}
\caption{\label{fig1} Novaco-McTague model for orientational ordering of (confined) monolayers. (a-c) Ball models of three stacked hexagonal lattices. The top (black) and bottom (blue) lattices are displayed only on the bottom and top halves, respectively. The lattice parameter $a$ increases (a-c) between the values for a monolayer locked to the bottom (a) and top (c) surfaces, and the layer's orientation $\phi$ deviates from 30$^\circ$ in (b). Moir\'{e} unit vectors are shown with black arrows. (b) Inset: zoom on a possible SDW (displacement field $\vec{u}$, black arrows). (d) Side-view cartoon of a graphene/Na/Ir stack. Electron donation from Na to C and Ir produces out-of-plane electrostatic dipoles, altogether in spring-like $u^2$ pairwise interactions; additionally Na atoms experience interaction potentials, with distinct periodicities, from Ir and graphene ($V^\mathrm{b=Ir}$, $V^\mathrm{t=C}$). (e,f) Simulated monolayer energy term $E(\phi,a)$ without (e) and with (f) a bottom Ir(111).}
\end{center}
\end{figure*}

\textit{\textbf{Modified Novaco-McTague model. -- }}Unlike in one-dimensional counterparts, the equilibrium states of (2D) monolayers onto periodic surfaces can have an ordered structure above 0~K. Model Hamiltonians unfortunately generally have no exact solution, especially when the units of the monolayer have a truly 2D positional degree of freedom \cite{braun}. Exploration of the corresponding phase diagram hence often relies on methods such as Monte Carlo \cite{schmidt,hamilton,salamacha} and molecular dynamics \cite{abraham,jin} simulations.

The ground state is a periodic 2D array of interfacial topological defects called dislocations, or, said differently, a 2D array of atomic in-plane (quasi)coincidences between the monolayer and substrate lattices \cite{braun} (Figs.~\ref{fig1}a-c). Commensurate arrays exist, whose unit vectors are related by integer-matrix transformations to the surface or monolayer structure \cite{doering,artaud}. These phases correspond to a locking of the monolayer onto the surface lattice (Figs.~\ref{fig1}a,c). Non-commensurate phases also exist, which depart from the locked phases by a deviation of $\phi$ and their lattice parameter $a$ (Fig.~\ref{fig1}b). A characteristic variation of $\phi$ \textit{versus} $a$ has been predicted by Novaco and McTague.

In the approximation of planar atomic displacements, the Hamiltonian of the confined monolayer comprises (\textit{i}) the energy of the undistorted monolayer, (\textit{ii}) the atoms' kinetic energy (adiabatic approximation, electrons in their ground states), (\textit{iii}) the elastic energy, and (\textit{iv}) the interaction energy between the two rigid surfaces and the monolayer's units \cite{mctague}:

\begin{eqnarray}
\hat{H} & = & E_0 + \sum_j \frac{1}{2M} \hat{p}_j\hat{p}_j + \frac{1}{2}\sum_{i,j}\Phi_{i,j}\hat{u}_i\hat{u}_j \nonumber \\
 & & + \sum_j \sum_{\vec{G}} ({V}^{\mathrm{b}}_{\vec{G}}+{V}^{\mathrm{t}}_{\vec{G}}) \mathrm{e}^{i\vec{G}\cdot\vec{R}_j} \mathrm{e}^{i\vec{G}\cdot\vec{u}_j} \label{Hamilton}
\end{eqnarray}

A dyadic representation with $x$ and $y$ cartesian indexes has been used (implying the corresponding sums), $j$ is the index of an atom of the monolayer, $\vec{u}_j$ its in-plane displacement with respect to the position $\vec{R}_{j}$ in the undistorted layer, $\hat{u}_j$ and $\hat{p}_j$ the displacement and momentum operators associated to the atom, and $\Phi_{i,j}$ the force constant matrix. In (\textit{iv}) the sum runs over all $\vec{G}$ vectors of reciprocal space, and two sets of Fourier components $V^{\mathrm{b}}_{\vec{G}}$ and $V^{\mathrm{t}}_{\vec{G}}$ describe the monolayer interaction with the bottom and top surface respectively. Following Fuselier, Raich and Gillis, the free energy derived from Eq.~\ref{Hamilton}\, comprises a negative term, $E$, translating a stabilization \textit{via} periodic local planar deformations forming a so-called static distorsion wave (SDW) \cite{fuselier}, which writes:

\begin{equation}
-\frac{N}{2} \sum_{\vec{G},\lambda} \phi_{\vec{G}} \vec{G} \cdot \vec{\epsilon}_{\lambda}(\vec{G}) \frac{1}{M \omega_{\lambda}^2(\vec{G})} \phi_{\vec{G}} \vec{G} \cdot \vec{\epsilon}_{\lambda}(\vec{G}) \label{E_SDW}
\end{equation}

Here, $\phi_{\vec{G}}$ is a Fourier component of $\phi_j = \sum_i V^{\mathrm{b}}(\vec{S}_i - \vec{R}_j) + \sum_i V^{\mathrm{t}}(\vec{S}_i - \vec{R}_j)$, with $\vec{S}_i$ a direct lattice vector of the bottom and top surfaces; $\vec{\epsilon}_{\lambda=1,2}$ are eigenvectors of frequency $\omega_{\lambda=1,2}$ corresponding to the two in-plane acoustic (longitudinal and transverse) phonon modes of the monolayer. In the sum, terms of higher weight have smaller $\omega_{\lambda}^2(\vec{G})$ values. They are low-frequency (long-wavelength) phonons, first transverse ones then longitudinal ones, associated to the SDW sketched in the inset of Fig.~\ref{fig1}b.

We now examine $E$ numerically, and for that purpose select a system that will be experimentally relevant, a monolayer of Na sandwiched between graphene and an Ir(111) surface. Sodium atoms (and alkali atoms in general) donate part of their electrons to graphene \cite{papagno} and presumably to Ir as well, and are thus associated with electric dipoles repelling each other (Fig.~\ref{fig1}d) \cite{SM}. The repulsive interactions in the lattice configuration add up to a harmonic-like (elastic) interaction within the monolayer \cite{over}. The Na-graphene and Na-Ir(111) interactions can each be described with six first-order Fourier components.

Figures~\ref{fig1}e,f show the calculated $E(\phi,a)$ in the cases of Na/graphene and graphene/Na/Ir(111). Two $\phi$ minima at $30^\circ\pm\phi_\mathrm{m}$ are observed for all $a$ values: the monolayer orientational orderings, clockwise and anticlockwise, are degenerate -- in practice the monolayer will break into domains twisted by $\pm\phi_\mathrm{m}$. While in the absence of the Ir(111) surface, $\phi_\mathrm{m}$ monotonously increases with $a$, this is not the case for the confined monolayer: $\phi_\mathrm{m}$ has a maximum as $a$ increases, meaning that a two-way twist, back and forth, occurs. This is the main new prediction derived from our modified Novaco-McTague model.

Why this is so can be intuited in terms of frustration arising from two competing interactions. The $(\phi_\mathrm{m},a)$ traces cross the configurations with the monolayer locked on the graphene or Ir(111) lattices [assuming $(\sqrt{3}\times\sqrt{3})R30$ superstructures named Ir$_3$Na and C$_6$Na, see Figs.~\ref{fig1}a,c]. Away from these configurations, the monolayer can be incommensurate and still orientational-ordered. Yet, the top and bottom surfaces influence the monolayer's structure, with commensurate energy terms (derived from Eq.~\ref{Hamilton}) that exceed $E$ only when $a/\sqrt{3}$ approaches graphene's and Ir(111)'s lattice constants ($a_\mathrm{graphene}$, $a_\mathrm{Ir}$). The influence of the surfaces is apparent in SDWs, an analysis of which is given in the Supplemental Material \cite{SM}.

\textit{\textbf{Practical realisation of confined monolayers. -- }}Both graphite \cite{wu} and dense-packed noble metal surfaces \cite{doering_b} allow the orientational ordering of alkali atom monolayers. In agreement with the original Novaco-McTague picture, this ordering is changing with the monolayer density, which is simply realised by varying the dose of alkali atoms on the surface \cite{wu,doering_b}. The observations are usually done below room temperature, while at higher temperature thermal excitation of defects and atomic desorption prevent any kind of stable or long-lived ordering.

Promising materials for observing the two-way twist could combine the surfaces of graphite and dense-packed metal surfaces. These surfaces have similar symmetry, which is valuable for a simple proof of concept, and typically differ by 10\% in their lattice constant. Single layers of graphite (graphene) grow on metal surfaces with a high quality, and we opted for graphene/Ir(111), which we prepared with a single orientation by chemical vapour deposition with ethylene under ultrahigh vacuum \cite{vangastel,SM}. Alkali metals often intercalate between graphene and the metal substrate \cite{gruneis,petrovic,andersen,halle,struzzi,daukiya}, including Na and Cs \cite{papagno,petrovic} (the former, at large enough doses \cite{pervan}) we chose for this reason. In our experiments, they were deposited using a resistively-heated source, taking the precautions needed to avoid contaminations of the highly reactive alkali atoms \cite{lindgren,*politano,SM}. The observations described below were made at room temperature.

\begin{figure}[!ht]
\begin{center}
\includegraphics[width=8cm]{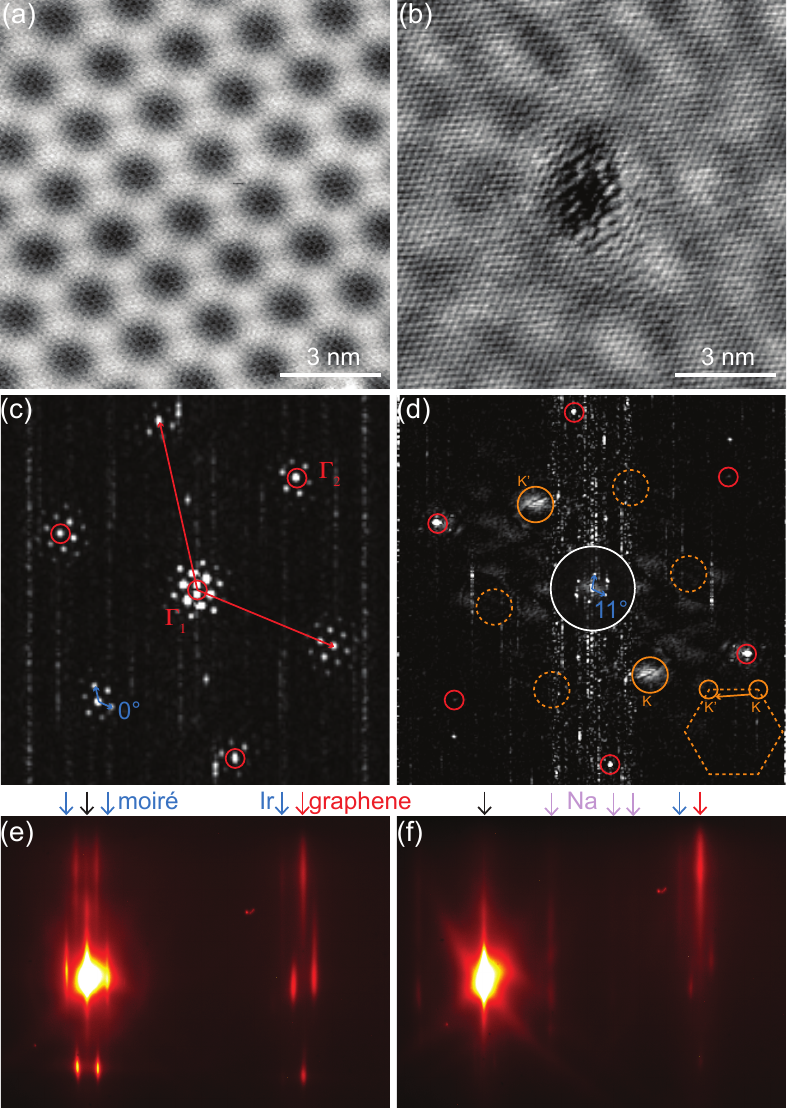}
\caption{\label{fig2} Experimental realisation of a confined monolayer. (a,b) STM topographs of graphene/Ir(111) (0.7~V, 2.20~nA) (a) and graphene/Na/Ir(111) (-0.1~V, 2.38~nA) with a Na coverage $\rho$ = 0.38$\pm$0.01$>1/3$, the density of Ir$_3$Na (b). (c,d) Fast Fourier transforms of (a,b) revealing the graphene harmonics (red vectors pointing to Brillouin zone centers $\Gamma_1$, $\Gamma_2$, red circles) and the moir\'{e} harmonics around them (blue vectors). In (d), the color scale in the central region has been adjusted to reveal the moir\'{e} harmonics, and an electronic interference pattern around the $K$ points is highlighted (orange solid circles). Inset: Sketch of graphene's Brillouin zone and electronic scattering events. (e,f) RHEED patterns ($[1\bar{2}1]$ azimuth, 17~keV) of graphene/Ir(111) (a) and graphene/Na/Ir(111) (b). Arrows mark the specular reflection (black), graphene (red), Ir(111) (blue), moir\'{e} (blue), and Na (violet) streaks.}
\end{center}
\end{figure}

Before and after Na dosing, we discerne graphene's atomic lattice but very different moir\'{e} patterns with STM (Figs.~\ref{fig2}a,b). As previously observed with Cs intercalation, the (apparent) height of the moir\'{e} nanoripples is much smaller \cite{halle_b}. Fourier transforms of the STM images reveal that the in-plane structure has also changed. While for pristine graphene/Ir(111) the moir\'{e} and graphene lattices have the same orientation (Fig.~\ref{fig2}c) as expected for a high quality graphene/Ir(111) \cite{vangastel}, after Na deposition they form a 11$^\circ$ angle (Fig.~\ref{fig2}d). Their unit vectors have lengths in a ratio increasing from 10.4 to 11.3. Since graphene's lattice constant hardly varies ($\sim$0.1\% according to RHEED) upon Na deposition, the latter means that the intercalated alkali layer gives access to larger moir\'{e} lattice constants than the maximum value of $\sim$ 2.5~nm characteristic of graphene/Ir(111). A geometrical transformation composed of a rotation of $\sim$ 2$^\circ$ and a compression of $\sim$ 4\% of the Na layer, with respect to an Ir$_3$Na commensurate structure, accounts for the observations in Figs.~\ref{fig2}b,d \cite{SM}. This gives a quantitative estimate of the density, $\rho$ = 0.38$\pm$0.01 [the atomic density of Ir(111) being $\rho$ = 1], consistent with the deposition rate assessed by RHEED \cite{SM}.

Figure~\ref{fig2}b purposely features one of the characteristic defects appearing after Na deposition. They are separated by several 10~nm, and could be pathways for Na intercalation under graphene. Strikingly, they are surrounded by lines arranged in a $(\sqrt{3}\times\sqrt{3})R30$ pattern with respect to graphene. They have a strong contribution in the Fourier transform (within solid orange circles in Fig.~\ref{fig2}d; the absence of signals in the doted orange circles is attributed to asymmetry of the STM tip or of the defect's structure). Their origin are electronic interferences between states in graphene occupying distinct valleys in the band structure (see inset of Fig.~\ref{fig2}d). While they are absent in graphene/Ir(111) due to a slight tendency to hybridisation between graphene and Ir(111)'s orbitals \cite{busse}, they are characteristic of systems where the substrate-graphene interaction is marginal or quenched \cite{rutter,mallet,yang,dombrowski}.

RHEED confirms that the initial moir\'{e} harmonics, found close to the center of reciprocal space, vanish upon Na intercalation (compare Figs.~\ref{fig2}e,f). In the meantime, new diffraction streaks appear (Fig.~\ref{fig2}f), whose positions are those expected for a Na layer slightly rotated and strained ($\sim$ 2$^\circ$ and 4\%, see above) with respect to a Ir$_3$Na superstructure. Similar observations are made with Cs.

The intercalated systems are remarkable for the room temperature ordering of the alkali layers. Ordering, otherwise occurring at lower temperatures, is here ascribed to a confinement-induced hinderance of thermal excitations and desorption.

\begin{figure}[!ht]
\begin{center}
\includegraphics[width=8cm]{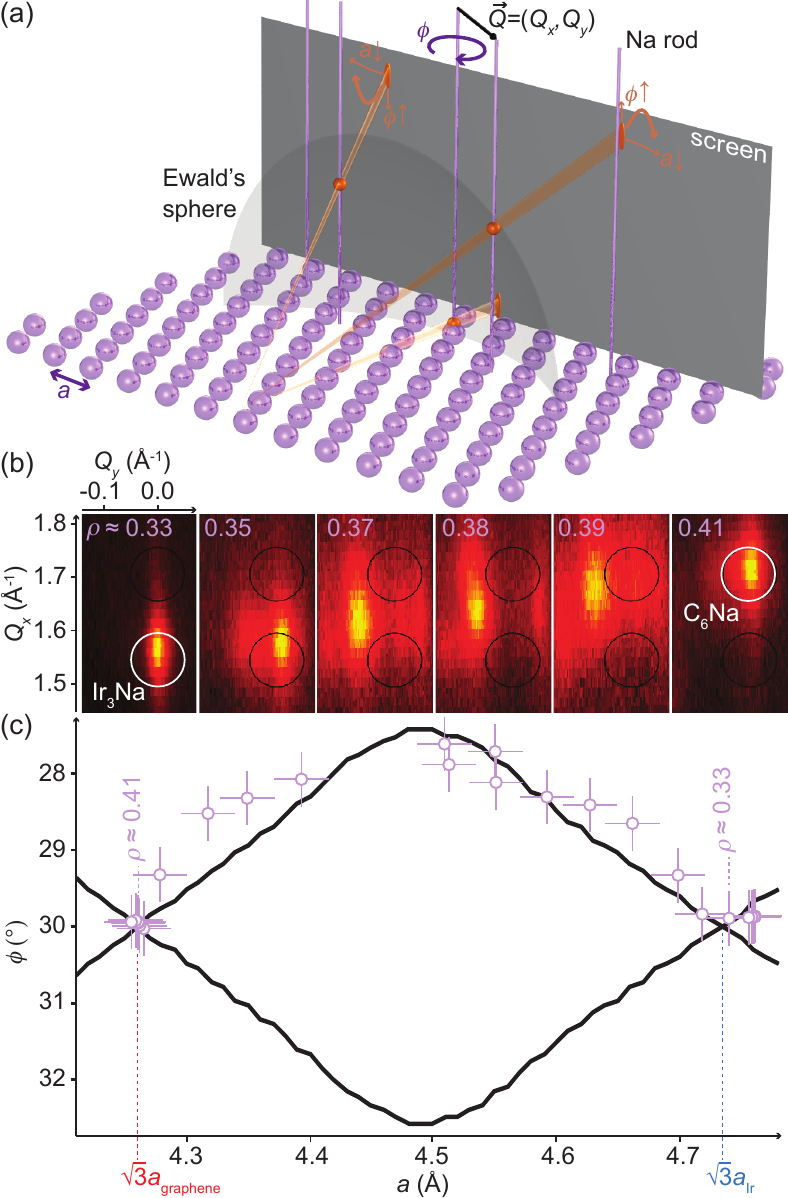}
\caption{\label{fig3} Experimental evidence of two-way orientational ordering. (a) RHEED geometry: the Na monolayer reciprocal space rods intersecting the Ewald's sphere define the diffraction angles. As the monolayer twists clockwise and the Na rods too, diffraction streaks move up (down) vertically at the left (right) side of reciprocal space. A planar compression moves the streaks horizontally. The trajectories predicted with the modified Novaco-McTague model are shown with curved orange arrows. (b) Monolayer streak in the ($Q_x$, $Q_y$) plane, at increasing Na densities, from the Ir$_3$Na Ir-locked ($\rho$=1/3) to the C$_6$Na graphene-locked ($\rho$$\simeq$0.41) phases. (c) $\phi$ and $a$ values (violet data points), extracted from the RHEED movies with 2D fits, as the Na density increases (0.005 increase between successive points). The black curve is the energy minimum calculated with our modified Novaco-McTague model (Fig.~\ref{fig1}f).}
\end{center}
\end{figure}

\textit{\textbf{Orientational ordering in alkali monolayers. -- }} Now, we unveil, with RHEED monitoring during alkali atom deposition, the influence of the intercalated layer density on its orientational ordering. As shown in Fig.~\ref{fig3}a and in the animation found in the Supplemental Material \cite{SM}, the diffraction streaks of the monolayer are expected to move as $\phi$ and $a$ vary. As it is well known, they move along the horizontal as $a$ varies. Less often made use of, the streaks move vertically as $\phi$ changes. Overall, a streak movement along the two directions originates from changes of both $a$ and $\phi$. Such a mixed movement (curved arrows in Fig.~\ref{fig3}a) is observed during Na and Cs evaporation, as shown in sequences of RHEED patterns focused on one of the alkali layer diffraction streaks \cite{SM}.

To track the variations of $a$ and $\phi$, a new coordinate system can be used to represent RHEED patterns (see animation in the Supplemental Material \cite{SM}), in the $(Q_x,Q_y)$ plane [$\vec{Q}=(Q_x,Q_y,Q_z)$ being the scattering vector], \textit{i.e.} much like low-energy electron diffraction patterns, instead of the usual $(Q_y,Q_z)$ representation.

Figure~\ref{fig3}b presents some of these patterns taken for increasing Na doses. The left and right patterns show peaks, corresponding to the Ir$_3$Na and C$_6$Na commensurate locked phases. Starting from the Ir$_3$Na phase, the monolayer diffraction spot follows a path that is not parallel to $Q_y$ or $Q_x$. This implies that the monolayer progressively rotates and compresses. Similar observations are made when changing the Na deposition rate and for Cs deposition \cite{SM}.

Figure~\ref{fig3}c translates these $(Q_x,Q_y)$ variations in terms of $(\phi,a)$ for the full RHEED sequence \cite{SM}. $\phi$ decreases by $\simeq$2.5$^\circ$ as $a$ \textit{decreases} by $\simeq$4\%, then the same compression leads to a $\phi$ \textit{increase} of 2.5$^\circ$, back to the $\phi=30^\circ$ value, corresponding now to a graphene-locked commensurate phase. The experimental $(\phi,a)$ points are overlaid onto the trace of the minima of $E$, calculated from the modified Novaco-McTague model (same simulation as in Fig.~\ref{fig1}f). The data match reasonably well the top-branch of the simulation (the bottom branch is not accessible experimentally, as it would show up below the horizon defined by the sample surface). Overall, it appears that orientational ordering indeed occurs within the confined Na and Cs monolayers, and that the Novaco-McTague picture for a succession of incommensurate phases reasonably accounts for this phenomenon.

Noteworthy, the diffracted intensity is minimum for $\phi\simeq30+2.5^\circ$ (see sequences of RHEED patterns \cite{SM}). The reason could be a proliferation ($\phi<30+2.5^\circ$) and healing ($\phi>30+2.5^\circ$) of stacking imperfections at the Ir(111)/alkali and alkali/graphene interfaces, or a progressive modification of the SDWs as $\phi$ departs from 0$^\circ$, acquiring local curls in their displacement field (Fig.~S1b in the Supplemental Material \cite{SM}), thereby altering diffraction intensities.

\textit{\textbf{Conclusions and prospects. -- }}We have adapted the model introduced by Novaco and McTague, to consider the competing interactions imposed onto an atomic monolayer by a top and a bottom periodic surface. Our model predicts a very characteristic orientational ordering, with a back-and-forth twist of the incommensurate monolayer and rather unique SDWs. We then tested our model with Na and Cs monolayers confined within the  narrow gap between graphene and Ir(111), and indeed found the sought for two-way twist as the monolayer's lattice parameter increases.

A direct extension of our work is the observation of the expectedly small-amplitude SDWs that our model predicts \cite{SM}, for instance by high resolution low temperature STM imaging (similarly to non-confined molecular monolayers \cite{meissner}), or by spot profile analysis of high resolution synchrotron X-ray surface scattering data. Much remains to explore about the effect of temperature, of the nature of the confined monolayer or confining walls (which could both be 2D materials), of lattice symmetries, and about the physical properties emerging from the peculiar orientational ordering and the competition of three kinds of orders, tunable \textit{via} the layer density.


%

\end{document}